\documentclass[11pt]{article}
\usepackage{amsfonts}
\usepackage{amsmath}

\newcommand{\be}{\begin{equation}}
\newcommand{\ee}{\end{equation}}
\newcommand{\bea}{\begin{eqnarray}}
\newcommand{\eea}{\end{eqnarray}}

\makeatletter

\@addtoreset{equation}{section}
\def\section{\@startsection {section}{1}{\z@}{-3.5ex plus -1ex minus
 -.2ex}{2.3ex plus .2ex}{\large\bf\centering}}
\def\subsection{\@startsection{subsection}{2}{\z@}{-3.25ex plus -1ex minus -.2ex}{1.5ex plus .2ex}{\bf}}
\def\subsubsection{\@startsection{subsubsection}{3}{\z@}{-3.25ex plus -1ex minus -.2ex}{1.5ex plus .2ex}{\sl}}
\makeatother
\begin{document}

\baselineskip 18pt \parindent 12pt \parskip 12pt

\begin{titlepage}

\begin{center}
{\Large {\bf Quasideterminant solutions of an integrable chiral model in two dimensions }}\\\vspace{1in} {\large Bushra Haider   \footnote{%
bushrahaider@hotmail.com}  and M. Hassan   \footnote{%
mhassan@physics.pu.edu.pk } }\vspace{0.15in}

{{\it Department of Physics, University of the Punjab,\\
Quaid-e-Azam Campus, Lahore-54590, Pakistan.}}\\

\end{center}
\vspace{1cm}

\begin{abstract}
The Darboux transformation is used to obtain multisoliton
solutions of the chiral model in two dimensions. The matrix
solutions of the principal chiral model and its Lax pair are
expressed in terms of quasideterminants. The iteration of Darboux
transformation gives the quasideterminant multisoliton solutions
of the model. It has been shown that the quasideterminant
multisoliton solution of the chiral model is the same as obtained
by Zakharov and Mikhailov using the dressing method based on
matrix Riemann-Hilbert problem.
\end{abstract}
\vspace{1cm} PACS: 11.10.Nx, 02.30.Ik
\end{titlepage} 

\section{Introduction}

\smallskip\ \ The principal chiral model (chiral field taking values in a
Lie group) is a well known example of integrable models of relativistically
invariant lagrangian field theories in two dimensions \cite{Novikov:1984id}-%
\cite{mikh}. The principal chiral models belong to more general family of
two dimensional integrable field theories, known as symmetric space sigma
models, where the fundamental fields take values in symmetric spaces as
their target spaces. The soliton solutions of various sigma models have been
obtained using the inverse scattering method and the multisoliton solutions
are obtained by means of Darboux-Backlund transformations \cite%
{Novikov:1984id}-\cite{Harnad:1983we}. In this paper we study the
Darboux transformation of principal chiral model based on some Lie
group and express their soliton solutions in terms of
quasideterminants. We show that the matrix solutions of the
principal chiral model and those of its associated Lax pair are
expressed in terms of quasideterminants. The Darboux
transformation also leads to the quasideterminant expressions of
the conserved currents of the chiral field. We also obtain the
quasideterminant multisoliton solutions of the chiral model from
the $K$ times iteration of the Darboux transformations and relate
the quasideterminant multisoliton solutions of the chiral field
with the well-known solutions of Zakharov and Mikhailov
\cite{zakha} obtained by the matrix Riemann-Hilbert problem. At
the end, we discuss the solution of the chiral model based on the
Lie group $SU\left( 2\right) $. We also study the asymptotic
behaviour of the solution.

The principal chiral field $g(x)$ with values in some Lie group $\mathcal{G}$
is governed by the Lagrangian\footnote{%
The spacetime conventions are such that the light-cone coordinates $x^{\pm }$
are related to the orthonormal coordinates by $x^{\pm }=\frac{1}{2}\left(
t\pm x\right) $ with the derivatives $\partial _{\pm }=\frac{1}{2}\left(
\partial _{t}\pm \partial _{x}\right) .$}
\begin{equation}
L=\frac{1}{2}\mbox{Tr}\left( \partial _{+}g^{-1}\partial _{-}g\right) ,
\label{action}
\end{equation}%
with $g^{-1}g=gg^{-1}=I$. The $\mathcal{G}$-valued field $g\left(
x^{+},x^{-}\right) $ can be expressed as
\begin{equation}
g\left( x^{+},x^{-}\right) \equiv e^{i\pi _{a}T^{a}}=1+i\pi _{a}T^{a}+\frac{1%
}{2}\left( i\pi _{a}T^{a}\right) ^{2}+\cdots ,  \label{field exp}
\end{equation}%
where $\pi _{a}$ is in the Lie algebra $\mathbf{g}$ of the Lie group $%
\mathcal{G}$ and $T^{a},a=1,2,3,\ldots ,\text{dim}\mathbf{g},$ are
anti-hermitian matrices with the normalization Tr$\left( T^{a}T^{b}\right)
=-\delta ^{ab}$ and are the generators of $\mathcal{G}$ in the fundamental
representation satisfying%
\begin{equation}
\left[ T^{a},T^{b}\right] =f^{abc}T^{c},  \label{generators}
\end{equation}%
where $f^{abc}$ are the structure constants of the Lie algebra $\mathbf{g}$.
For any $X\in \mathbf{g}$, we write $X=X^{a}T^{a}$ and $X^{a}=-$Tr$\left(
T^{a}X\right) .$ The action (\ref{action}) is invariant under a global
continuous symmetry%
\begin{equation}
{\mathcal{G}_{L}}\times {\mathcal{G}_{R}}:\mbox{ \ \ \ \ \ \ \ \ \ \ \ }%
g\left( x^{+},x^{-}\right) \longmapsto UgV^{-1},  \label{gsymmetry}
\end{equation}%
where $U\in \mathcal{G}_{L}$ and $V\in \mathcal{G}_{R}.$ The Noether
conserved current associated with the $\mathcal{G}_{R}$ transformation is $%
j_{\pm }=-g^{-1}\partial _{\pm }g$, which takes values in the Lie algebra $%
\mathbf{g},$ so that one can decompose the current into components $j_{\pm
}\left( x^{+},x^{-}\right) =j_{\pm }^{a}\left( x^{+},x^{-}\right) T^{a}.$
The conserved current corresponding to the $\mathcal{G}_{L}$ transformation
is $-gj_{\pm }g^{-1}$. The equation of motion of the principal chiral model
is the conservation equation and the zero curvature condition
\begin{eqnarray}
\partial _{+}j_{-}+\partial _{-}j_{+} &=&0,  \label{EOM} \\
\partial _{-}j_{+}-\partial _{+}j_{-}+[j_{+},j_{-}] &=&0.
\label{zc condition}
\end{eqnarray}%
The equations of motion (\ref{EOM})-(\ref{zc condition}) can be written as
the compatibility condition of the following Lax pair
\begin{eqnarray}
\ {\partial }_{+}V({\lambda }) &=&\frac{1}{1-\lambda }j_{+}V({\lambda }%
)\noindent ,\   \label{V11} \\
\ {\partial }_{-}V({\lambda }) &=&\frac{1}{1+\lambda }\ j_{-}V({\lambda }%
)\noindent \ ,\ \   \label{V12}
\end{eqnarray}%
where ${\lambda }$ is a real (or complex) parameter and $V$ is an invertible
$N\times N$ matrix, in general. We solve the Lax pair to find the matrix
solution $V(\lambda )$ such that $V(0)=g$. If we have any collection $%
(V(\lambda ),j_{\pm })$ which solves the Lax pair (\ref{V11})-(\ref{V12}),
then $V(0)=g$ solves the chiral field equation (\ref{EOM}). In the next
section, we define the Darboux transformation via a Darboux matrix on matrix
solutions $V$ of the Lax pair (\ref{V11})-(\ref{V12}). To write down the
explicit expressions for matrix solutions of the chiral model, we will use
the notion of quasideterminant introduced by Gelfand and Retakh \cite{GR}-%
\cite{krob}.

Let $X$ be an $N\times N$ matrix over a ring $R$ (noncommutative, in
general). For any $1\leq i$, $j\leq N$, let $r_{i}$ be the $i$th row and $%
c_{j}$ be the $j$th column of $X$. There exist $N^{2}$ quasideterminants
denoted by $|X|_{ij}$ for $i,j=1,\ldots ,N$ and are defined by
\begin{equation}
|X|_{ij}=\left\vert
\begin{array}{cc}
X^{ij} & c_{j}^{\,\,i} \\
r_{i}^{\,\,j} & \frame{\fbox{$x_{ij}$}}%
\end{array}%
\right\vert =x_{ij}-r_{i}^{\,\,j}\left( X^{ij}\right) ^{-1}c_{j}^{\,\,i},
\label{quasid}
\end{equation}%
where $x_{ij}$ is the $ij$th entry of $X$, $r_{i}^{\,\,j}$ represents the $i$%
th row of $X$ without the $j$th entry, $c_{j}^{\,\,i}$ represents the $j$th
column of $X$ without the $i$th entry and $X^{ij}$ is the submatrix of $X$
obtained by removing from $X$ the $i$th row and the $j$th column. The
quasideterminats are also denoted by the following notation. If the ring $R$
is commutative i.e. the entries of the matrix $X$ all commute, then
\begin{equation}
|X|_{ij}=(-1)^{i+j}\frac{\mathrm{det}X}{\mathrm{det}X^{ij}}.
\end{equation}%
For a detailed account of quasideterminants and their properties see e.g.
\cite{GR}-\cite{krob}. In this paper, we will consider only
quasideterminants that are expanded about an $N\times N$ matrix over a
commutative ring. Let
\begin{equation}
\left(
\begin{array}{cc}
A & B \\
C & D%
\end{array}%
\right) ,  \notag
\end{equation}%
be a block decomposition of any $K\times K$ matrix where the matrix $D$ is $%
N\times N$ and $A$ is invertible. The ring $R$ in this case is the
(noncommutative) ring of $N\times N$ matrices over another commutative ring.
The quasideterminant of $K\times K$ matrix expanded about the $N\times N$
matrix $D$ is defined by
\begin{equation}
\left\vert
\begin{array}{cc}
A & B \\
C & \frame{\fbox{$D$}}%
\end{array}%
\right\vert =D-CA^{-1}B.
\end{equation}%
An important property of quasideterminants is the noncommutative Jacobi
identity. For a general quasideterminant expanded about an $N\times N$
matrix $D$, we have
\begin{equation}
\left\vert
\begin{array}{ccc}
E & F & G \\
H & A & B \\
J & C & \frame{\fbox{$D$}}%
\end{array}%
\right\vert =\left\vert
\begin{array}{cc}
E & G \\
J & \frame{\fbox{$D$}}%
\end{array}%
\right\vert -\left\vert
\begin{array}{cc}
E & F \\
J & \frame{\fbox{$C$}}%
\end{array}%
\right\vert \left\vert
\begin{array}{cc}
E & F \\
H & \frame{\fbox{$A$}}%
\end{array}%
\right\vert ^{-1}\left\vert
\begin{array}{cc}
E & G \\
H & \frame{\fbox{$B$}}%
\end{array}%
\right\vert .  \label{NCJ}
\end{equation}%
From the noncommutative Jacobi identity, we get the homological relation%
\begin{equation}
\left\vert
\begin{array}{ccc}
E & F & G \\
H & A & \frame{\fbox{$B$}} \\
J & C & D%
\end{array}%
\right\vert =\left\vert
\begin{array}{ccc}
E & F & O \\
H & A & \frame{\fbox{$O$}} \\
J & C & I%
\end{array}%
\right\vert \left\vert
\begin{array}{ccc}
E & F & G \\
H & A & B \\
J & C & \frame{\fbox{$D$}}%
\end{array}%
\right\vert ,  \label{HR}
\end{equation}%
where $O$ and $I$ denote the null and identity matrices respectively. The
quasideterminants have found various applications in the theory of
integrable systems, where the multisoliton solutions of various
noncommutative integrable systems are expressed in terms of
quisideterminants (see e.g. \cite{Nimmo}-\cite{Hamanaka1}).

\section{Darboux transformation}

The Darboux transformation is one of the well known methods of obtaining
multisoliton solutions of integrable systems \cite{darboux}-\cite{matveev}.
We define the Darboux transformation on the matrix solutions of the Lax pair
(\ref{V11})-(\ref{V12}), in terms of an $N\times N$ matrix $%
D(x^{+},x^{-},\lambda )$, called the Darboux matrix. For a general
discussion on Darboux matrix approach see e.g. \cite{sakh}-\cite{qing ji}.
The Darboux matrix relates the two matrix solutions of the Lax pair (\ref%
{V11})-(\ref{V12}), in such a way that the Lax pair is covariant under the
Darboux transformation. The Darboux transformation on the matrix solution of
the Lax pair (\ref{V11})-(\ref{V12}) is defined by
\begin{equation}
\widetilde{V}(\lambda )=D(x^{+},x^{-},{\lambda })V(\lambda )\ .  \label{Vd}
\end{equation}%
For the Lax pair (\ref{V11})-(\ref{V12}) to be covariant under the Darboux
transformation (\ref{Vd}), we require%
\begin{eqnarray}
\ {\partial }_{+}\widetilde{V}(\lambda ) &=&\frac{1}{1-\lambda }\widetilde{j}%
_{+}\widetilde{V}(\lambda ),  \label{v31} \\
\ {\partial }_{-}\widetilde{V}(\lambda ) &=&\frac{1}{1+\lambda }\ \widetilde{%
j}_{-}\widetilde{V}(\lambda )\ .  \label{v51}
\end{eqnarray}%
By substituting equation (\ref{Vd}) in equations (\ref{v31})-(\ref{v51}), we
get the following condition on the Darboux matrix $D(\lambda )$%
\begin{equation}
{\partial }_{\pm }D\left( \lambda \right) V\left( \lambda \right) +D\left(
\lambda \right) \frac{1}{1\mp \lambda }j_{\pm }V({\lambda })=\frac{1}{1\mp
\lambda }\widetilde{j}_{\pm }D\left( \lambda \right) V\left( \lambda \right)
.  \label{VD1}
\end{equation}%
For our system, we make the following ansatz for the Darboux matrix
\begin{equation}
D(x^{+},x^{-},\lambda )=\lambda I-S(x^{+},x^{-}),\ \   \label{darbouxm1}
\end{equation}%
where $S(x^{+},x^{-})$ is some $N\times N$ matrix to be determined and $I$
is an $N\times N$ identity matrix. Note that we consider here the Darboux
matrix of degree one which is linear in $\lambda $. Therefore, to construct
the Darboux matrix $D(x^{+},x^{-},\lambda )$, it is only necessary to
determine the matrix $S(x^{+},x^{-})$. Now substituting (\ref{Vd}) in
equation (\ref{VD1}) and using (\ref{V11})-(\ref{V12}), we get the following
Darboux transformation for the Lie algebra valued conserved currents $%
\widetilde{j}_{\pm }$
\begin{eqnarray}
\widetilde{j}_{+} &=&j_{+}+\ {\partial }_{+}S,  \notag \\
\ \ \ \widetilde{j}_{-} &=&\ j_{-}-\ \partial _{-}S,\   \label{abts}
\end{eqnarray}%
and the matrix $S$ is subjected to satisfy the following conditions
\begin{eqnarray}
\partial _{+}S(I-S) &=&[j_{+},S],\   \label{sas} \\
\partial _{-}S(I+S) &=&[j_{-},S]\ .  \label{scs}
\end{eqnarray}%
These new transformed currents are also conserved and curvature free i.e.
\begin{eqnarray}
\partial _{+}\widetilde{j}_{-}+\partial _{-}\widetilde{j}_{+} &=&0, \\
\partial _{-}\widetilde{j}_{+}-\partial _{+}\widetilde{j}_{-}+[\widetilde{j}%
_{+},\widetilde{j}_{-}] &=&0.
\end{eqnarray}%
Now we proceed to determine the matrix $S$, so that the explicit Darboux
transformation in terms of particular solutions of the Lax pair, can be
constructed.

Let ${\lambda }_{1},\cdots ,{\lambda }_{N},$ be $N$ distinct real (or
complex) constant parameters and ${\lambda }_{i}\neq \pm 1;i=1,2,\cdots ,N.$
Let us also define $N$ constant column vectors $\left\vert 1\right\rangle
,\left\vert 2\right\rangle ,\cdots ,\left\vert N\right\rangle $, such that
\begin{equation}
M=\left( V({\lambda }_{1})\left\vert 1\right\rangle ,\cdots ,V({\lambda }%
_{N})\left\vert N\right\rangle \right) =\left( \left\vert m_{1}\right\rangle
,\cdots ,\left\vert m_{N}\right\rangle \right) ,  \label{HS}
\end{equation}%
be an invertible $N\times N$ matrix. Each column $\left\vert
m_{i}\right\rangle =V({\lambda }_{i})\left\vert i\right\rangle $ in $M$ is a
column solution of the Lax pair (\ref{V11})-(\ref{V12}) when ${\lambda }={%
\lambda }_{i}.$ i.e., it satisfies
\begin{eqnarray}
\ {\partial }_{+}\left\vert m_{i}\right\rangle &=&\frac{1}{1-\lambda _{i}}%
j_{+}\left\vert m_{i}\right\rangle \ ,  \label{hV3} \\
\ {\partial }_{-}\left\vert m_{i}\right\rangle &=&\frac{1}{1+\lambda _{i}}\
j_{-}\left\vert m_{i}\right\rangle \ ,  \label{hV5}
\end{eqnarray}%
and $i=1,2,\ldots ,N$. If we define an $N\times N$ matrix of particular
eigenvalues as
\begin{equation}
\Lambda =\text{diag}({\lambda }_{1},\ldots ,{\lambda }_{N}),  \label{la}
\end{equation}%
then the Lax pair (\ref{hV3}) and (\ref{hV5}) can be written in $N\times N$
matrix form as%
\begin{eqnarray}
\ {\partial }_{+}M &=&j_{+}M\left( I-\Lambda \ \right) ^{-1},  \label{hhV3}
\\
\ {\partial }_{-}M &=&j_{-}M\left( I+\Lambda \ \right) ^{-1},  \label{hhV5}
\end{eqnarray}%
where the $N\times N$ matrix $M$ is a particular matrix solution of the Lax
pair (\ref{V11})-(\ref{V12}) with $\Lambda $ being a matrix of particular
eigenvalues. In terms of particular matrix solution $M$ of the Lax pair (\ref%
{V11})-(\ref{V12}), we define the matrix $S$ to be
\begin{equation}
S=M\Lambda M^{-1}\text{.}  \label{so}
\end{equation}%
Now we show that the matrix $S$ defined in (\ref{so}), satisfies equations (%
\ref{sas})-(\ref{scs}). First, we take the $x^{+}$ derivative of the matrix (%
\ref{so}), so that we have
\begin{eqnarray}
\partial _{+}S &=&\partial _{+}(M\Lambda M^{-1}),  \notag \\
&=&\partial _{+}M\Lambda M^{-1}+M\Lambda \partial _{+}(M^{-1}),  \notag \\
&=&j_{+}M(I-\Lambda )^{-1}\Lambda M^{-1}-M\Lambda M^{-1}j_{+}M(I-\Lambda
)^{-1}M^{-1},  \notag \\
&=&-j_{+}+M(I-\Lambda )M^{-1}j_{+}M(I-\Lambda )^{-1}M^{-1},  \notag \\
&=&-j_{+}+\left( I-S\right) j_{+}\left( I-S\right) ^{-1},  \label{ds}
\end{eqnarray}%
which is the equation (\ref{sas}). Similarly, operating $\partial _{-}$ on (%
\ref{so}), we get
\begin{equation}
\partial _{-}S=j_{-}-\left( I+S\right) j_{-}\left( I+S\right) ^{-1},
\label{ds1}
\end{equation}%
which is nothing but the equation (\ref{scs}). This shows that the choice (%
\ref{so}) of the matrix $S$ satisfies all the conditions imposed by the
covariance of the Lax pair under the Darboux transformation. Therefore, we
say that the transformation
\begin{eqnarray}
\widetilde{V} &=&(\lambda I-M\Lambda M^{-1})V,  \notag \\
\widetilde{j}_{\pm } &=&M(I\mp \Lambda )M^{-1}j_{\pm }M(I\mp \Lambda
)^{-1}M^{-1},  \label{dd}
\end{eqnarray}%
is the required Darboux transformation of the chiral model in terms of
particular matrix solution $M$ with the particular eigenvalue matrix $%
\Lambda $. Let us now introduce a primitive field $F_{\pm }$ such
that $j_{\pm }=F_{\pm }F_{\pm }^{-1}$, which transforms in a
simple way under the Darboux transformation i.e, %
\begin{eqnarray}
\tilde{F}_{\pm } &=&M(I\mp \Lambda )M^{-1}F_{\pm }.  \label{dd1}
\end{eqnarray}%
The Darboux transformation on the chiral field $g(x)$ is now
defined by
\begin{equation}
\widetilde{g}\,=\,\widetilde{V}(0)\,=\,-\left( M\Lambda M^{-1}\right) g.
\label{g1}
\end{equation}%
Since we have assumed $M$ to be invertible therefore, we require that det$%
M\neq 0$. At this stage, we conclude that if the collection $(V,j_{\pm })$
is a solution of the Lax pair (\ref{V11})-(\ref{V12}) and the matrix $S$ is
defined by (\ref{so}), then $(\widetilde{V},\widetilde{j}_{\pm })$ defined
by (\ref{dd}) by means of Darboux transformation (\ref{darbouxm1}), is also
a solution of the same Lax pair. This establishes the covariance of the Lax
pair (\ref{V11})-(\ref{V12}) under the Darboux transformation (\ref%
{darbouxm1}).

If the chiral fields take values in the Lie group $U(N)$, then we also
require for the new solutions to take values in $U(N)$. We know that the Lie
group $U(N)$ consists of all $N\times N$ matrices $g$ such that $g^{\dagger
}=g^{-1}$. An arbitrary matrix $X$ belongs to the Lie algebra $\mathbf{u}(N)$
of the Lie group $U(N)$ if and only if $X^{\dagger }=-X$. Since the currents
$j_{\pm }$ are $\mathbf{u}(N)$ valued, therefore, we require that the new
currents $\widetilde{j}_{\pm }$ obtained by Darboux transformation must be $%
\mathbf{u}(N)$ valued i.e. they must be anti-hermitian. This leads to the
following condition on the matrix $S$:
\begin{equation}
\partial _{\pm }\left( S+S^{\dagger }\right) =0.  \label{ah}
\end{equation}%
For the matrix $S$ to satisfy (\ref{ah}), we proceed by taking specific
values of parameters ${\lambda }_{i}$. Let $\mu $ be a non-zero complex
number and ${\lambda }_{i}=\mu \,\,\,\,(i=1,2,\dots N)$. Now choose $%
\left\vert i\right\rangle $ such that
\begin{equation}
\left\langle m_{i}\right. \left\vert m_{j}\right\rangle =0\quad \quad
\mathrm{for}\quad {\lambda }_{i}\neq {\lambda }_{j},  \label{horthogonal}
\end{equation}%
holds everywhere and $\left\vert m_{i}\right\rangle $ are all linearly
independent. From the definition of the matrix $S$ it can be observed that
\begin{equation}
\left\langle m_{i}\right\vert \left( S^{\dagger }+S\right) \left\vert
m_{j}\right\rangle =\left( \bar{{\lambda }}_{i}+{\lambda }_{j}\right)
\left\langle m_{i}\right. \left\vert m_{j}\right\rangle ,  \label{ss3}
\end{equation}%
implying that $\left\langle m_{i}\right. \left\vert m_{j}\right\rangle =0$,
when ${\lambda }_{i}\neq {\lambda }_{j}$. If ${\lambda }_{i}={\lambda }%
_{j}=\mu \,\,$, we have
\begin{equation}
\left\langle m_{i}\right\vert \left( S^{\dagger }+S\right) \left\vert
m_{j}\right\rangle =\left\langle m_{i}\right\vert \left( \mu +\bar{\mu}%
\right) \left\vert m_{j}\right\rangle .  \label{ss4}
\end{equation}%
Since $\left\vert m_{i}\right\rangle $'s are all linearly independent,
therefore, equation (\ref{ss4}) implies
\begin{equation}
\left( S^{\dagger }+S\right) =\left( \mu +\bar{\mu}\right) I,  \label{sssum}
\end{equation}%
which further implies (\ref{ah}). Again from the Lax pair (\ref{V11})-(\ref%
{V12}), we have%
\begin{equation*}
\left\langle m_{i}\right\vert S^{\dagger }S\left\vert m_{j}\right\rangle
=\left\langle m_{i}\right\vert \bar{{\lambda }}_{i}{\lambda }_{j}\left\vert
m_{j}\right\rangle ,
\end{equation*}%
thus, if ${\lambda }_{i}=\mu $,%
\begin{equation}
S^{\dagger }S=\bar{\mu}\mu .  \label{ssproduct}
\end{equation}%
For the Lie group $SU(N)$, we have to impose further condition on the new
conserved currents $\widetilde{j}_{\pm }$. We know that the Lie group $SU(N)$
consists of all $N\times N$ matrices $g$ such that $g\in U(N)$ and $\mathrm{%
det}g=1$. An arbitrary matrix $X$ belongs to the Lie algebra $\mathbf{su}(N)$
of the group $SU(N)$, if and only if $\mathrm{Tr}X=0$. So if the chiral
field $g(x)$ take values in $SU(N)$, then we also require that $\mathrm{Tr}%
j_{\pm }=0$, $\mathrm{Tr}\widetilde{j}_{\pm }=0$; and for this to be the
case, the matrix $S$ is required to satisfy
\begin{equation}
\mathrm{Tr}\partial _{\pm }S=0.  \label{ss5}
\end{equation}%
The condition $\mathrm{Tr}\partial _{\pm }S=0$ is satisfied by equations (%
\ref{ds})-(\ref{ds1}), using the cyclicity of trace.

We impose the reality condition on solutions $V(\lambda )$ of the Lax pair (%
\ref{V11})-(\ref{V12})
\begin{equation}
V^{\dagger }\left( \bar{\lambda}\right) V\left( \lambda \right) =V^{\dagger
}\left( \bar{\lambda}\right) V\left( \lambda \right) \in \mathrm{Span}%
\left\{ I\right\} .  \label{reality}
\end{equation}%
where $I$ is an $N\times N$ unit matrix and $\mathrm{Span}\{I\}$ is the
subspace of the underlying Lie group spanned by $I$. To obtain well-defined
transformed solutions, the Darboux transformation must preserve this reality
condition i.e.,%
\begin{equation}
\widetilde{V}^{\dagger }\left( \bar{\lambda}\right) \widetilde{V}\left(
\lambda \right) \in \mathrm{Span}\left\{ I\right\} .  \label{reality1}
\end{equation}%
Using (\ref{Vd}) and (\ref{darbouxm1}), also making use of (\ref{ssproduct}%
),(\ref{sssum}) we see that
\begin{equation*}
\widetilde{V}^{\dagger }\left( \bar{\lambda}\right) \widetilde{V}\left(
\lambda \right) =\left( \lambda ^{2}-\lambda \left( \mu +\bar{\mu}\right)
+\mu \bar{\mu}\right) V^{\dagger }\left( \bar{\lambda}\right) V\left(
\lambda \right) \in \mathrm{Span}\left\{ I\right\} ,
\end{equation*}%
i.e. the transformed solution satisfies the reality condition, or in other
words, the Darboux transformation preserves the reality condition (\ref%
{reality1}). In the next section, we will express the solutions of chiral
model obtained by the Darboux transformation, in terms of quasideterminants
that are expanded about an $N\times N$ matrix over a commutative ring.
\section{Quasideterminant solutions}
Since the particular solution $M$ of the Lax pair (\ref{V11})-(\ref{V12}) is
an invertible $N\times N$ matrix, therefore we can express the Darboux
transformations (\ref{dd}) and (\ref{g1}) in terms of quasideterminants. The
Darboux transformed solution $\widetilde{V}$ of the Lax pair (\ref{V11})-(%
\ref{V12}) is expressed as
\begin{eqnarray}
\widetilde{V} &=&\left( \lambda I-S\right) V\,=\,\left( \lambda I-M\Lambda
M^{-1}\right) V,  \notag \\
&=&\left\vert
\begin{array}{cc}
M & I \\
M\Lambda & \frame{\fbox{$\lambda I$}}%
\end{array}%
\right\vert V,  \label{V1qu}
\end{eqnarray}%
and the chiral field $\widetilde{g}$ is expressed as
\begin{equation}
\widetilde{g}=\widetilde{V}(0)\,=\,-Sg=-\left( M\Lambda M^{-1}\right)
g\,=\,\left\vert
\begin{array}{cc}
M & I \\
M\Lambda & \frame{\fbox{$O$}}%
\end{array}%
\right\vert g.
\end{equation}%
Similarly from (\ref{dd1}) the conserved currents $\tilde{j}_{\pm }$ are
expressed as
\begin{equation}
\widetilde{j}_{\pm }=\tilde{F}_{\pm }\tilde{F}_{\pm }^{-1}=\left\vert
\begin{array}{cc}
M & I \\
M\left( I\mp \Lambda \right) & \frame{\fbox{$O$}}%
\end{array}%
\right\vert j_{\pm }\left\vert
\begin{array}{cc}
M & I \\
M\left( I\mp \Lambda \right) & \frame{\fbox{$O$}}%
\end{array}%
\right\vert ^{-1}.  \label{JQ1}
\end{equation}%
For the next iteration of Darboux transformation, we take $M_{1}$, $M_{2}$
be two particular solutions of the Lax pair (\ref{hhV3})-(\ref{hhV5}) at $%
\Lambda =\Lambda _{1}$ and $\Lambda =\Lambda _{2}$ respectively. Using the
notation $V\left[ 1\right] =V,g\left[ 1\right] =g,j_{\pm }\left[ 1\right]
=j_{\pm },F_{\pm }\left[ 1\right] =F_{\pm }$ and $V\left[ 2\right] =%
\widetilde{V},g\left[ 2\right] =\widetilde{g},j_{\pm }\left[ 2\right] =%
\widetilde{j}_{\pm },F_{\pm }\left[ 2\right] =\tilde{F}_{\pm }$, we write
two-fold Darboux transformation on $V$ as%
\begin{equation*}
V\left[ 3\right] =\left( \lambda I-S\left[ 2\right] \right) \left( \lambda
I-S\left[ 1\right] \right) V=\left( \lambda I-S\left[ 2\right] \right) V%
\left[ 2\right] ,
\end{equation*}%
where $S\left[ 1\right] =M_{1}\Lambda _{1}M_{1}^{-1}.$ By writing $S\left[ 2%
\right] =M\left[ 2\right] \Lambda _{2}M\left[ 2\right] ^{-1},$ we get%
\begin{equation}
V\left[ 3\right] =\left( \lambda I-M\left[ 2\right] \Lambda _{2}M\left[ 2%
\right] ^{-1}\right) V\left[ 2\right] ,  \label{Vit3}
\end{equation}%
where $M\left[ 2\right] =\left. V\left[ 2\right] \right\vert _{V\rightarrow
M_{2}}$, so that after the action of $\lambda I-S\left[ 1\right] ,$ the
vector $\left\vert m_{j}^{(2)}\right\rangle $ transforms as $\left( \lambda
_{j}^{(2)}I-S\left[ 1\right] \right) \left\vert m_{j}^{\left( 2\right)
}\right\rangle $. Therefore, we have%
\begin{equation}
M\left[ 2\right] =\left( M_{2}\Lambda _{2}-S\left[ 1\right] M_{2}\right)
=\left\vert
\begin{array}{cc}
M_{1} & M_{2} \\
M_{1}\Lambda _{1} & \frame{\fbox{$M_{2}\Lambda _{2}$}}%
\end{array}%
\right\vert .  \label{M12}
\end{equation}%
By using equation (\ref{V1qu}) and (\ref{M12}) in (\ref{Vit3}), we get%
\begin{eqnarray}
V\left[ 3\right] &=&\lambda \left\vert
\begin{array}{cc}
M_{1} & I \\
M_{1}\Lambda _{1} & \frame{\fbox{$\lambda I$}}%
\end{array}%
\right\vert V-\left\vert
\begin{array}{cc}
M_{1} & M_{2} \\
M_{1}\Lambda _{1} & \frame{\fbox{$M_{2}\Lambda _{2}$}}%
\end{array}%
\right\vert \Lambda _{2}\left\vert
\begin{array}{cc}
M_{1} & M_{2} \\
M_{1}\Lambda _{1} & \frame{\fbox{$M_{2}\Lambda _{2}$}}%
\end{array}%
\right\vert ^{-1}\left\vert
\begin{array}{cc}
M_{1} & I \\
M_{1}\Lambda _{1} & \frame{\fbox{$\lambda I$}}%
\end{array}%
\right\vert V,  \notag \\
&=&\left\vert
\begin{array}{cc}
M_{1}\Lambda _{1} & \lambda V \\
M_{1}\Lambda _{1}^{2} & \frame{\fbox{$\lambda ^{2}V$}}%
\end{array}%
\right\vert -\left\vert
\begin{array}{cc}
M_{1}\Lambda _{1} & M_{2}\Lambda _{2} \\
M_{1}\Lambda _{1}^{2} & \frame{\fbox{$M_{2}\Lambda _{2}^{2}$}}%
\end{array}%
\right\vert \left\vert
\begin{array}{cc}
M_{1}\Lambda _{1} & M_{2}\Lambda _{2} \\
M_{1} & \frame{\fbox{$M_{2}$}}%
\end{array}%
\right\vert ^{-1}\left\vert
\begin{array}{cc}
M_{1}\Lambda _{1} & \lambda V \\
M_{1} & \frame{\fbox{$V$}}%
\end{array}%
\right\vert ,  \notag \\
&=&\left\vert
\begin{array}{ccc}
M_{1} & M_{2} & I \\
M_{1}\Lambda _{1} & M_{2}\Lambda _{2} & \lambda I \\
M_{1}\Lambda _{1}^{2} & M_{2}\Lambda _{2}^{2} & \frame{\fbox{$\lambda ^{2}I$}%
}%
\end{array}%
\right\vert V,  \label{V311}
\end{eqnarray}%
where we have used homological relation (\ref{HR}) in the second step and
the noncommutative Jacobi identity (\ref{NCJ}) in the last step.

Similarly the two-fold Darboux transformation on conserved currents $j_{\pm
} $ gives
\begin{equation}
j_{\pm }\left[ 3\right] =F_{\pm }\left[ 3\right] F_{\pm }\left[ 3\right]
^{-1},\text{ }  \label{QJQ}
\end{equation}%
where the factor $F_{\pm }\left[ 3\right] $ is expressed in terms of
quasideterminants as%
\begin{eqnarray*}
F_{\pm }\left[ 3\right] &=&\left( M\left[ 2\right] \left( I\mp \Lambda
_{2}\right) M\left[ 2\right] ^{-1}\right) \left( M\left[ 1\right] \left(
I\mp \Lambda _{1}\right) M\left[ 1\right] ^{-1}\right) F_{\pm } \\
&=&-\left\vert
\begin{array}{cc}
M_{1} & M_{2} \\
M_{1}\Lambda _{1} & \frame{\fbox{$M_{2}\Lambda _{2}$}}%
\end{array}%
\right\vert \left( I\mp \Lambda _{2}\right) \left\vert
\begin{array}{cc}
M_{1} & M_{2} \\
M_{1}\Lambda _{1} & \frame{\fbox{$M_{2}\Lambda _{2}$}}%
\end{array}%
\right\vert ^{-1}\left\vert
\begin{array}{cc}
M_{1} & I \\
M_{1}\left( I\mp \Lambda _{1}\right) & \frame{\fbox{$O$}}%
\end{array}%
\right\vert F_{\pm } \\
&=&-\left\vert
\begin{array}{cc}
M_{1} & M_{2} \\
M_{1}\left( I\mp \Lambda _{1}\right) & \frame{\fbox{$M_{2}\left( I\mp
\Lambda _{2}\right) $}}%
\end{array}%
\right\vert \left( I\mp \Lambda _{2}\right) \\
&&\times \left\vert
\begin{array}{cc}
M_{1} & M_{2} \\
M_{1}\left( I\mp \Lambda _{1}\right) & \frame{\fbox{$M_{2}\left( I\mp
\Lambda _{2}\right) $}}%
\end{array}%
\right\vert ^{-1}\left\vert
\begin{array}{cc}
M_{1} & I \\
M_{1}\left( I\mp \Lambda _{1}\right) & \frame{\fbox{$O$}}%
\end{array}%
\right\vert F_{\pm } \\
&=&-\left\vert
\begin{array}{cc}
M_{1} & M_{2}\left( I\mp \Lambda _{2}\right) \\
M_{1}\left( I\mp \Lambda _{1}\right) & \frame{\fbox{$M_{2}\left( I\mp
\Lambda _{2}\right) ^{2}$}}%
\end{array}%
\right\vert \left\vert
\begin{array}{cc}
M_{1} & M_{2} \\
M_{1}\left( I\mp \Lambda _{1}\right) & \frame{\fbox{$M_{2}\left( I\mp
\Lambda _{2}\right) $}}%
\end{array}%
\right\vert ^{-1}\left\vert
\begin{array}{cc}
M_{1} & I \\
M_{1}\left( I\mp \Lambda _{1}\right) & \frame{\fbox{$O$}}%
\end{array}%
\right\vert F_{\pm } \\
&=&\left\vert
\begin{array}{ccc}
M_{1} & M_{2} & I \\
M_{1}\left( I\mp \Lambda _{1}\right) & M_{2}\left( I\mp \Lambda _{2}\right)
& O \\
M_{1}\left( I\mp \Lambda _{1}\right) ^{2} & M_{2}\left( I\mp \Lambda
_{2}\right) ^{2} & \frame{\fbox{$O$}}%
\end{array}%
\right\vert F_{\pm },
\end{eqnarray*}%
where we have used the homological relation (\ref{HR}) and noncommutative
Jacobi identity (\ref{NCJ}) in obtaining the last step.

We can iterate the Darboux transformation $K$ times and obtain the
quasideterminant multisoliton solution of the chiral model. For each $%
k=1,2,\cdots ,K$, let $M_{k}$ be an invertible $N\times N$ matrix solution
of the Lax pair (\ref{V11})-(\ref{V12}) at $\Lambda =\Lambda _{k}$, then the
$K$-th solution $V[K+1]$ is expressed as
\begin{eqnarray}
V\left[ K+1\right]  &=&\prod\limits_{k=1}^{K}\left( \lambda I-S\left[ K-k+1%
\right] \right) V\,=\,\prod\limits_{k=1}^{K}\left\vert
\begin{array}{cc}
M\left[ K-k+1\right]  & I \\
M\left[ K-k+1\right] \Lambda _{K-k+1} & \frame{\fbox{$\lambda I$}}%
\end{array}%
\right\vert V,  \notag \\
&=&\lambda V\left[ K\right] -M\left[ K\right] \Lambda _{K}M\left[ K\right]
^{-1}V\left[ K\right] ,  \notag \\
&=&\left\vert
\begin{array}{ccccc}
M_{1} & M_{2} & \cdots  & M_{K} & I \\
M_{1}\Lambda _{1} & M_{2}\Lambda _{2} & \cdots  & M_{K}\Lambda _{K} &
\lambda I \\
M_{1}\Lambda _{1}^{2} & M_{2}\Lambda _{2}^{2} & \cdots  & M_{K}\Lambda
_{K}^{2} & \lambda ^{2}I \\
\vdots  & \vdots  & \cdots  & \vdots  & \vdots  \\
M_{1}\Lambda _{1}^{K} & M_{2}\Lambda _{2}^{K} & \cdots  & M_{K}\Lambda
_{K}^{K} & \frame{\fbox{$\lambda ^{K}I$}}%
\end{array}%
\right\vert V.  \label{VNqu}
\end{eqnarray}%
The above results can be proved by induction using the properties of
quasideterminants. First we see that the result (\ref{VNqu}) is true for $K=1
$ and gives equation (\ref{V1qu}) directly. Next we consider%
\begin{eqnarray}
V\left[ K+2\right]  &=&\left( \lambda I-S\left[ K+1\right] \right) V\left[
K+1\right] ,  \notag \\
&=&\lambda V\left[ K+1\right] -S\left[ K+1\right] V\left[ K+1\right] ,
\notag \\
&=&\lambda V\left[ K+1\right] -M\left[ K+1\right] \Lambda _{K+1}M\left[ K+1%
\right] ^{-1}V\left[ K+1\right] .  \label{VK2}
\end{eqnarray}%
By using equation (\ref{VNqu}) in the expression (\ref{VK2}) and using the
fact that $M\left[ i\right] =\left. V\left[ i\right] \right\vert
_{V\rightarrow M_{i}},$ we get%
\begin{eqnarray*}
V\left[ K+2\right]  &=&\left\vert
\begin{array}{ccccc}
M_{1} & M_{2} & \cdots  & M_{K} & \lambda I \\
M_{1}\Lambda _{1} & M_{2}\Lambda _{2} & \cdots  & M_{K}\Lambda _{K} &
\lambda ^{2}I \\
\vdots  & \vdots  & \cdots  & \vdots  & \vdots  \\
M_{1}\Lambda _{1}^{K} & M_{2}\Lambda _{2}^{K} & \cdots  & M_{K}\Lambda
_{K}^{K} & \frame{\fbox{$\lambda ^{K}I$}}%
\end{array}%
\right\vert V \\
&&-\left\vert
\begin{array}{ccccc}
M_{1} & M_{2} & \cdots  & M_{K} & M_{K+1} \\
M_{1}\Lambda _{1} & M_{2}\Lambda _{2} & \cdots  & M_{K}\Lambda _{K} &
M_{K+1}\Lambda _{K+1} \\
\vdots  & \vdots  & \cdots  & \vdots  & \vdots  \\
M_{1}\Lambda _{1}^{K} & M_{2}\Lambda _{2}^{K} & \cdots  & M_{K}\Lambda
_{K}^{K} & \frame{\fbox{$M_{K+1}\Lambda _{K+1}^{K}$}}%
\end{array}%
\right\vert \Lambda _{K+1} \\
&&\times \left\vert
\begin{array}{ccccc}
M_{1} & M_{2} & \cdots  & M_{K} & M_{K+1} \\
M_{1}\Lambda _{1} & M_{2}\Lambda _{2} & \cdots  & M_{K}\Lambda _{K} &
M_{K+1}\Lambda _{K+1} \\
\vdots  & \vdots  & \cdots  & \vdots  & \vdots  \\
M_{1}\Lambda _{1}^{K} & M_{2}\Lambda _{2}^{K} & \cdots  & M_{K}\Lambda
_{K}^{K} & \frame{\fbox{$M_{K+1}\Lambda _{K+1}^{K}$}}%
\end{array}%
\right\vert ^{-1} \\
&&\left\vert
\begin{array}{ccccc}
M_{1} & M_{2} & \cdots  & M_{K} & \lambda I \\
M_{1}\Lambda _{1} & M_{2}\Lambda _{2} & \cdots  & M_{K}\Lambda _{K} &
\lambda ^{2}I \\
\vdots  & \vdots  & \cdots  & \vdots  & \vdots  \\
M_{1}\Lambda _{1}^{K} & M_{2}\Lambda _{2}^{K} & \cdots  & M_{K}\Lambda
_{K}^{K} & \frame{\fbox{$\lambda ^{K}I$}}%
\end{array}%
\right\vert V.
\end{eqnarray*}%
Now rearranging the above expression and using the noncommutative Jacobi
identity (\ref{NCJ}) and homological relations (\ref{HR}), we get%
\begin{eqnarray}
V\left[ K+2\right]  &=&\left\vert
\begin{array}{ccccc}
M_{1}\Lambda _{1} & M_{2}\Lambda _{2} & \cdots  & M_{K}\Lambda _{K} &
\lambda V \\
M_{1}\Lambda _{1}^{2} & M_{2}\Lambda _{2}^{2} & \cdots  & M_{K}\Lambda
_{K}^{2} & \lambda ^{2}V \\
\vdots  & \vdots  & \cdots  & \vdots  & \vdots  \\
M_{1}\Lambda _{1}^{K+1} & M_{2}\Lambda _{2}^{K+1} & \cdots  & M_{K}\Lambda
_{K}^{K+1} & \frame{\fbox{$\lambda ^{K}V$}}%
\end{array}%
\right\vert   \notag \\
&&-\left\vert
\begin{array}{ccccc}
M_{1}\Lambda _{1} & M_{2}\Lambda _{2} & \cdots  & M_{K}\Lambda _{K} &
M_{K+1}\Lambda _{K+1} \\
M_{1}\Lambda _{1}^{2} & M_{2}\Lambda _{2}^{2} & \cdots  & M_{K}\Lambda
_{K}^{2} & M_{K+1}\Lambda _{K+1}^{2} \\
\vdots  & \vdots  & \cdots  & \vdots  & \vdots  \\
M_{1}\Lambda _{1}^{K+1} & M_{2}\Lambda _{2}^{K+1} & \cdots  & M_{K}\Lambda
_{K}^{K+1} & \frame{\fbox{$M_{K+1}\Lambda _{K+1}^{K+1}$}}%
\end{array}%
\right\vert   \notag \\
&&\times \left\vert
\begin{array}{ccccc}
M_{1}\Lambda _{1} & M_{2}\Lambda _{2} & \cdots  & M_{K}\Lambda _{K} &
M_{K+1}\Lambda _{K+1} \\
M_{1}\Lambda _{1}^{2} & M_{2}\Lambda _{2}^{2} & \cdots  & M_{K}\Lambda
_{K}^{2} & M_{K+1}\Lambda _{K+1}^{2} \\
\vdots  & \vdots  & \cdots  & \vdots  & \vdots  \\
M_{1}\Lambda _{1}^{K} & M_{2}\Lambda _{2}^{K} & \cdots  & M_{K}\Lambda
_{K}^{K} & M_{K+1}\Lambda _{K+1}^{K} \\
M_{1} & M_{2} & \cdots  & M_{K} & \frame{\fbox{$M_{K+1}$}}%
\end{array}%
\right\vert ^{-1}  \notag \\
&&\times \left\vert
\begin{array}{ccccc}
M_{1}\Lambda _{1} & M_{2}\Lambda _{2} & \cdots  & M_{K}\Lambda _{K} &
\lambda V \\
M_{1}\Lambda _{1} & M_{2}\Lambda _{2} & \cdots  & M_{K}\Lambda _{K} &
\lambda ^{2}V \\
\vdots  & \vdots  & \cdots  & \vdots  & \vdots  \\
M_{1}\Lambda _{1}^{K} & M_{2}\Lambda _{2}^{K} & \cdots  & M_{K}\Lambda
_{K}^{K} & \lambda ^{K}V \\
M_{1} & M_{2} & \cdots  & M_{K} & \frame{\fbox{$V$}}%
\end{array}%
\right\vert ,  \notag \\
&=&\left\vert
\begin{array}{ccccc}
M_{1} & M_{2} & \cdots  & M_{K+1} & I \\
M_{1}\Lambda _{1} & M_{2}\Lambda _{2} & \cdots  & M_{K+1}\Lambda _{K+1} &
\lambda I \\
M_{1}\Lambda _{1}^{2} & M_{2}\Lambda _{2}^{2} & \cdots  & M_{K+1}\Lambda
_{K+1}^{2} & \lambda ^{2}I \\
\vdots  & \vdots  & \cdots  & \vdots  & \vdots  \\
M_{1}\Lambda _{1}^{K+1} & M_{2}\Lambda _{2}^{K+1} & \cdots  & M_{K+1}\Lambda
_{K+1}^{K+1} & \frame{\fbox{$\lambda ^{K+1}I$}}%
\end{array}%
\right\vert V.  \label{VK2J}
\end{eqnarray}%
Therefore (\ref{VNqu}) is verified. The multisoliton solution $g[K+1]$ of
the chiral model can be readily obtained by taking $\lambda =0$ in the
expression of $V[K+1]$ i.e.
\begin{eqnarray}
g\left[ K+1\right]  &=&\prod\limits_{k=1}^{K}\left( -1\right) ^{k}S\left[
K-k+1\right] g\,=\,\prod\limits_{k=1}^{K}\left\vert
\begin{array}{cc}
M\left[ K-k+1\right]  & I \\
M\left[ K-k+1\right] \Lambda _{K-k+1} & \frame{\fbox{$O$}}%
\end{array}%
\right\vert g,  \notag \\
&=&\left\vert
\begin{array}{ccccc}
M_{1} & M_{2} & \cdots  & M_{K} & I \\
M_{1}\Lambda _{1} & M_{2}\Lambda _{2} & \cdots  & M_{K}\Lambda _{K} & O \\
M_{1}\Lambda _{1}^{2} & M_{2}\Lambda _{2}^{2} & \cdots  & M_{K}\Lambda
_{K}^{2} & O \\
\vdots  & \vdots  & \cdots  & \vdots  & \vdots  \\
M_{1}\Lambda _{1}^{K} & M_{2}\Lambda _{2}^{K} & \cdots  & M_{K}\Lambda
_{K}^{K} & \frame{\fbox{$O$}}%
\end{array}%
\right\vert g.  \label{gk3}
\end{eqnarray}%
Similarly the $K$ times iteration of Darboux transformation gives the
following expression of the conserved currents
\begin{equation}
j_{\pm }\left[ K+1\right] =F_{\pm }\left[ K+1\right] F_{\pm }\left[ K+1%
\right] ^{-1},  \label{JK}
\end{equation}%
where%
\begin{eqnarray}
F_{\pm }\left[ K+1\right]  &=&\left( I\mp S\left[ K\right] \right) \cdots
\left( I\mp S\left[ 2\right] \right) \left( I\mp S\left[ 1\right] \right)
F_{\pm },  \notag \\
&&\prod\limits_{k=1}^{K}\left\vert
\begin{array}{cc}
M\left[ K-k+1\right]  & I \\
M\left[ K-k+1\right] \left( I\mp \Lambda _{K-k+1}\right)  & \frame{\fbox{$O$}%
}%
\end{array}%
\right\vert F_{\pm },  \label{FK1} \\
&=&\left\vert
\begin{array}{ccccc}
M_{1} & M_{2} & \cdots  & M_{K} & I \\
M_{1}\left( I\mp \Lambda _{1}\right)  & M_{2}\left( I\mp \Lambda _{2}\right)
& \cdots  & M_{K}\left( I\mp \Lambda _{K}\right)  & O \\
M_{1}\left( I\mp \Lambda _{1}\right) ^{2} & M_{2}\left( I\mp \Lambda
_{2}\right) ^{2} & \cdots  & M_{K}\left( I\mp \Lambda _{K}\right) ^{2} & O
\\
\vdots  & \vdots  & \cdots  & \vdots  & \vdots  \\
M_{1}\left( I\mp \Lambda _{1}\right) ^{K} & M_{2}\left( I\mp \Lambda
_{2}\right) ^{K} & \cdots  & M_{K}\left( I\mp \Lambda _{K}\right) ^{K} &
\frame{\fbox{$O$}}%
\end{array}%
\right\vert F_{\pm }.  \label{FK}
\end{eqnarray}%
The expression (\ref{FK}) can also be proved by induction in the same way as we did for (\ref{VNqu}). Hence we see that the equations (\ref{FK1}) and (\ref{FK}) together with (%
\ref{JK}) are the required expressions of $K$-th conserved
currents of the chiral model expressed in terms of
quasideterminants involving particular solutions of the linear
problem associated with the chiral model. Note that the equations
(\ref{VNqu}) and (\ref{gk3}) are the required quasideterminant
expressions for the $K$th iteration of $V$ and $g$.

\section{Relation with Zakharov-Mikhailov's dressing method}

In this section, we relate the quasideterminant mutisoliton solutions of the
previous section, with the solutions obtained by Zakharov and Mikhailov
using an equivalent method known as the dressing method. In the original
approach of Zakharov and Mikhailov, the analytical properties of the matrix
function $\mathcal{D}(\lambda )$ (now referred to as dressing function) are
studied in the complex $\lambda $-plane. In fact the dressing function $%
\mathcal{D}\left( \lambda \right) $ in Zakharov-Mikhailov approach is
equivalent to $\left( \lambda -\mu \right) ^{-1}D\left( \lambda \right) $,
where $D\left( \lambda \right) $ is the Darboux matrix (\ref{darbouxm1}).
Note that the equation (\ref{VD1}) remains invariant, if the
Darboux-dressing matrix is multiplied by a scalar factor. In particular, it
is required that $\mathcal{D}\left( \lambda \right) $ should be meromorphic
and $\mathcal{D}\left( \lambda \right) \rightarrow I$ as $\lambda
\rightarrow \pm \infty $. In other words, we say that the matrix function $%
\mathcal{D}\left( \lambda \right) $ has a pole at some $\lambda $ or any
entry of $\mathcal{D}\left( \lambda \right) $ has a pole at that particular
value of $\lambda $. If we take the simple case, in which $\mathcal{D}\left(
\lambda \right) $ has a single pole at $\lambda =\mu $, so that the dressing
function $\mathcal{D}\left( \lambda \right) $ is expressed in terms of a
hermition projector $P$. In what follows, we show that our Darboux matrix
expressed as a quasideterminant, can be written in terms of hermitian
projection operator, resulting in a solution of the chiral model without
much use of analytical properties of matrix functions involved.

The Darboux matrix (\ref{darbouxm1}) can also be written in terms of the
projector. For this purpose we make use of equation (\ref{ss3}) i.e., we
write
\begin{eqnarray*}
S\left\vert m_{i}\right\rangle &=&\lambda _{i}\left\vert m_{i}\right\rangle ,%
\text{ \ \ \ \ \ \ \ \ \ \ \ \ \ }i=1,2,\cdots ,n \\
S\left\vert m_{j}\right\rangle &=&\bar{\lambda}_{j}\left\vert
m_{j}\right\rangle ,\text{ \ \ \ \ \ \ \ \ \ \ \ \ }j=n+1,n+2,\cdots ,N.
\end{eqnarray*}%
Now we set $\lambda _{i}=\mu $ and $\lambda _{j}=\bar{\mu}$, so that the
matrix $S$ may be written as
\begin{equation}
S=\mu P+\bar{\mu}P^{\bot },  \label{SP}
\end{equation}%
where $P$ is the hermitian projector i.e., $P^{\dag }=P.$ Also we have $%
P^{2}=P$ and $P^{\bot }=I-P.$ The projector $P$ is completely characterized
by two subspaces $U=\mathrm{Im}P$ and $W=\mathrm{Ker}P$ given by the
condition $P^{\bot }U=0$ and $PW=0$, so that $P$ is defined as a hermitian
projection on a complex subspace and $P^{\bot }=I-P$ as projection on the
orthogonal space. Now the matrix $S$ is expressed as
\begin{eqnarray}
S &=&\mu P+\bar{\mu}(I-P),  \notag \\
&=&\left( \mu -\bar{\mu}\right) P+\bar{\mu}I.  \label{SP1}
\end{eqnarray}%
The Darboux matrix which is expressed as a quasideterminant in the previous
section can now be written as
\begin{eqnarray}
D\left( \lambda \right) &=&\left\vert
\begin{array}{cc}
M & I \\
M\Lambda & \frame{\fbox{$\lambda I$}}%
\end{array}%
\right\vert  \notag \\
&=&\lambda I-\left( \mu -\bar{\mu}\right) P-\bar{\mu}I,  \notag \\
&=&\left( \lambda -\bar{\mu}\right) I-\left( \mu -\bar{\mu}\right) P,  \notag
\\
&=&\left( \lambda -\bar{\mu}\right) \left( I-\frac{\mu -\bar{\mu}}{\lambda -%
\bar{\mu}}P\right) .  \label{DP}
\end{eqnarray}%
In the expression (\ref{DP}) the Darboux-dressing function, expressed as
quasideterminant containing the particular matrix solution $M$ of the Lax
pair \ref{V11})-(\ref{V12}), is shown to be expressed in terms of a
hermitian projector $P$ defined in terms of the particular column solutions $%
\left\vert m_{i}\right\rangle $ of the Lax pair. The Darboux-dressing
function (\ref{DP}) can also be used to obtain the multisoliton solution of
the system. For the case of models based on Lie groups of $N\times N$
matrices, we set $\lambda _{i}=\mu ,\left( i=1,2,\cdots ,n\right) $ and $%
\lambda _{j}=\bar{\mu},\left( j=n+1,\cdots ,N\right) $, so that the solution
$V[2]$ is given by
\begin{eqnarray*}
V\left[ 2\right] &=&\left( \lambda I-\mu \sum\limits_{i=1}^{n}\frac{%
\left\vert m_{i}\right\rangle \left\langle m_{i}\right\vert }{\left\langle
m_{i}\right\vert \left. m_{i}\right\rangle }-\bar{\mu}\sum\limits_{j=n+1}^{N}%
\frac{\left\vert m_{j}\right\rangle \left\langle m_{j}\right\vert }{%
\left\langle m_{j}\right\vert \left. m_{j}\right\rangle }\right) V, \\
&=&\left( \lambda -\bar{\mu}\right) \left( I-\frac{\mu -\bar{\mu}}{\lambda -%
\bar{\mu}}\sum\limits_{i=1}^{n}\frac{\left\vert m_{i}\right\rangle
\left\langle m_{i}\right\vert }{\left\langle m_{i}\right\vert \left.
m_{i}\right\rangle }\right) V\,=\,\left( \lambda -\bar{\mu}\right) \left( I-%
\frac{\mu -\bar{\mu}}{\lambda -\bar{\mu}}P\right) V.
\end{eqnarray*}%
The $K$-th time iteration then gives the $K$-th solution $V[K+1]$ of the Lax
pair
\begin{eqnarray}
V\left[ K+1\right] &=&\prod\limits_{k=1}^{K}\left\vert
\begin{array}{cc}
M\left[ K-k+1\right] & I \\
M\left[ K-k+1\right] \Lambda _{K-k+1} & \frame{\fbox{$\lambda I$}}%
\end{array}%
\right\vert V,  \notag \\
&=&\prod\limits_{k=1}^{K}\left( \lambda -\bar{\mu}_{K-k+1}\right) \left( I-%
\frac{\mu _{K-k+1}-\bar{\mu}_{K-k+1}}{\lambda -\bar{\mu}_{K-k+1}}P\left[
K-k+1\right] \right) V,  \label{VK}
\end{eqnarray}%
with the $(K+1)$-th soliton solution $g[K+1]$ of the chiral model given by
\begin{equation}
g\left[ K+1\right] =\prod\limits_{k=1}^{K}\left( -\bar{\mu}_{K-k+1}\right)
\left( I+\frac{\mu _{K-k+1}-\bar{\mu}_{K-k+1}}{\bar{\mu}_{K-k+1}}P\left[
K-k+1\right] \right) V,  \label{gk1}
\end{equation}%
where the hermitian projection in this case is
\begin{equation*}
P\left[ k\right] =\sum_{i=1}^{n}\frac{\left\vert m_{i}\left[ k\right]
\right\rangle \left\langle m_{i}\left[ k\right] \right\vert }{\left\langle
m_{i}\left[ k\right] \right\vert \left. m_{i}\left[ k\right] \right\rangle }%
,\quad \quad \quad k=1,2,\cdots ,K,
\end{equation*}%
with%
\begin{equation*}
\left\vert m_{i}\left[ k\right] \right\rangle =\left( \lambda _{i}^{\left(
k\right) }I-S\left[ k-1\right] \right) \left\vert m_{i}^{\left( k\right)
}\right\rangle ,
\end{equation*}%
and the $k$-th particular matrix solution $M_{k}$ of the Lax pair is written
in terms of $k$-th particular column solutions as
\begin{equation}
M_{k}=\left( \left\vert m_{1}^{\left( k\right) }\right\rangle ,\left\vert
m_{2}^{\left( k\right) }\right\rangle ,\cdots \left\vert m_{N}^{\left(
k\right) }\right\rangle \right) .
\end{equation}%
Now the expressions for the transformed currents $j_{\pm }\left[ K+1\right] $
are given by
\begin{equation}
j_{\pm }\left[ K+1\right] =\prod\limits_{k=1}^{K}\left( I\mp \frac{\left(
\mu _{K-k+1}-\bar{\mu}_{K-k+1}\right) }{\left( 1\mp \bar{\mu}_{K-k+1}\right)
}P\left[ K-k+1\right] \right) j_{\pm }\prod\limits_{l=1}^{K}\left( I\mp
\frac{\left( \bar{\mu}_{l}-\mu _{l}\right) }{\left( 1\mp \mu _{l}\right) }P%
\left[ l\right] \right) .  \label{jk1}
\end{equation}

The expressions (\ref{VK}), (\ref{gk1}) and (\ref{jk1}) can also be written
as sum of $K$ terms by using the condition that $V\left[ K\right] =0$ if $%
\lambda =\mu _{i},V=\left\vert m_{i}\right\rangle .$ The method is
illustrated in \cite{Novikov:1984id} (also see \cite{U(N)} for reference)$.$
The final expressions for $V\left[ K+1\right] ,g\left[ K+1\right] $ and $%
j_{\pm }\left[ K+1\right] $ are then given as%
\begin{eqnarray}
V\left[ K+1\right] &=&\sum\limits_{k,l=1}^{K}\left( \lambda -\bar{\mu}%
_{k}\right) \left( I-\frac{R_{k}}{\lambda -\bar{\mu}_{k}}\right) V,
\label{Vsum} \\
g\left[ K\right] &=&\sum\limits_{k,l=1}^{K}(-\bar{\mu}_{k})\left( I+\frac{%
R_{k}}{\bar{\mu}_{k}}\right) g,  \label{gsum} \\
j_{\pm }\left[ K\right] &=&\sum\limits_{k=1}^{K}\left( I\mp \frac{R_{k}}{%
1\mp \bar{\mu}_{k}}\right) j_{\pm }\sum\limits_{l=1}^{K}\left( I\mp \frac{%
R_{l}}{1\mp \mu _{l}}\right) .  \label{Jsum}
\end{eqnarray}%
where the function $R_{k}$ is defined by
\begin{equation}
R_{k}=\sum\limits_{l=1}^{K}\left( \mu _{l}-\bar{\mu}_{k}\right)
\sum_{i=1}^{n}\frac{\left\vert m_{i}^{\left( k\right) }\right\rangle
\left\langle m_{i}^{\left( l\right) }\right\vert }{\left\langle
m_{i}^{\left( k\right) }\right\vert \left. m_{i}^{\left( l\right)
}\right\rangle }
\end{equation}%
By expanding the right hand side in equation (\ref{Vsum}) and using (\ref%
{horthogonal}), we see that the two expressions for the $K$-th iteration of $%
V$, i.e. equations (\ref{VNqu}) and (\ref{Vsum}) are equivalent.

\section{The SU(2) model} In this
section we briefly discuss how to calculate the soliton solution of the
principal chiral model based on the Lie group $SU(2)$ using the method
outlined in the previous section. For the $SU(2)$ case, the solution has
been obtained in \cite{zakha}. Let us first calculate one-soliton solution
of the chiral model using the dressing (Darboux) matrix (\ref{DP}). The
matrix solution $V[1]$ of the Lax pair (\ref{V11})-(\ref{V12}) is given by
\begin{equation}
\widetilde{V}=\left( \lambda I-M\Lambda M^{-1}\right) V.  \label{p1}
\end{equation}%
Now for $N=2$ case, the particular solution $M$ of the Lax pair (\ref{V11})-(%
\ref{V12}) is given by an invertible $2\times 2$ matrix expressed in terms
of column solutions $\left\vert m_{1}\right\rangle $ and $\left\vert
m_{2}\right\rangle $: $M=\left(
\begin{array}{cc}
\left\vert m_{1}\right\rangle & \left\vert m_{2}\right\rangle%
\end{array}%
\right) $. We take the $2\times 2$ eigenvalue matrix $\Lambda $ to be $%
\Lambda =\left(
\begin{array}{cc}
\mu & 0 \\
0 & \bar{\mu}%
\end{array}%
\right) ,$ where we have taken $\lambda _{1}=\mu $ and $\lambda _{2}=\bar{\mu%
}$. With this the solution $V[2]$ is written as
\begin{eqnarray*}
\widetilde{V} &=&\left\vert
\begin{array}{cc}
M & I \\
M\Lambda & \frame{\fbox{$\lambda I$}}%
\end{array}%
\right\vert V\,=\,\left( \lambda I-\mu \frac{\left\vert m_{1}\right\rangle
\left\langle m_{1}\right\vert }{\left\langle m_{1}\right\vert \left.
m_{1}\right\rangle }+\bar{\mu}\frac{\left\vert m_{2}\right\rangle
\left\langle m_{2}\right\vert }{\left\langle m_{2}\right\vert \left.
m_{2}\right\rangle }\right) V, \\
&=&\left( \lambda I-\mu P-\bar{\mu}P^{\perp }\right) V\,=\,\left( \lambda -%
\bar{\mu}\right) \left( I-\frac{\mu -\bar{\mu}}{\lambda -\bar{\mu}}P\right) V
\end{eqnarray*}%
where the hermitian projection is $P=\frac{\left\vert m_{1}\right\rangle
\left\langle m_{1}\right\vert }{\left\langle m_{1}\right\vert \left.
m_{1}\right\rangle },$ with the orthogonal projection $P^{\perp }=I-P=\frac{%
\left\vert m_{2}\right\rangle \left\langle m_{2}\right\vert }{\left\langle
m_{2}\right\vert \left. m_{2}\right\rangle }.$ The Darboux matrix $D(\lambda
)$ as a quasideterminant may be expressed in terms of hermitian projection
and orthogonal projection as
\begin{equation}
D(\lambda )=\left\vert
\begin{array}{cc}
M & I \\
M\Lambda & \frame{\fbox{$\lambda I$}}%
\end{array}%
\right\vert =\left( \lambda -\bar{\mu}\right) \left( I-\frac{\mu -\bar{\mu}}{%
\lambda -\bar{\mu}}P\right) =\left( \lambda -\bar{\mu}\right) \left(
P^{\perp }+\frac{\lambda -{\mu }}{\lambda -\bar{\mu}}P\right) .
\end{equation}%
The one-soliton solution $\widetilde{g}$ in this case is given by
\begin{equation}
\widetilde{g}=\left\vert
\begin{array}{cc}
M & I \\
M\Lambda & \frame{\fbox{$\lambda I$}}%
\end{array}%
\right\vert \,g=-\bar{\mu}\left( I+\frac{\mu +\bar{\mu}}{\bar{\mu}}P\right)
\,g=-\bar{\mu}\left( P^{\perp }+\frac{{\mu }}{\bar{\mu}}P\right) .
\end{equation}%
For the construction of explicit solution in matrix form using the Darboux
transformation, we take the example of $\mathcal{G}=SU(2)$. The solutions
can be obtained by Darboux transformation by taking the trivial solution as
the seed solution. We have been considering the case where $j_{\pm }\in
\mathbf{su}(2)$, the following discussions, however, are essentially the
same for the Lie algebra $\mathbf{u}\left( 2\right) $. Let us take a most
general unimodular $2\times 2$ matrix representing an element of the Lie
algebra $\mathbf{su}\left( 2\right) $
\begin{equation*}
\ \left(
\begin{array}{ll}
X & Y \\
-\bar{Y} & \bar{X}%
\end{array}%
\right) ,
\end{equation*}%
where $Y$ and $X$ are complex numbers satisfying $X\bar{X}+Y\bar{Y}=1.$ Let $%
j_{\pm }$ be the non-zero constant (commuting) elements of $\mathbf{su}%
\left( 2\right) ,$ such that they are represented by anti-hermitian $2\times
2$ matrices
\begin{equation}
j_{+}=\left(
\begin{array}{ll}
{\mbox{i}}p & 0 \\
0 & -{\mbox{i}}p%
\end{array}%
\right) \ \ ,\ \ \ \ \ \ \ \ \ \ \ \ \ \ \ \ \ \ \ \ \ \ \ \ \ \ \ \ \ \
j_{-}=\left(
\begin{array}{ll}
{\mbox{i}}q & 0 \\
0 & -{\mbox{i}}q%
\end{array}%
\right) ,  \label{abab}
\end{equation}%
where $p,q$ are non-zero real numbers. The seed solution is then written as
\begin{equation}
\ g(x^{+},x^{-})=\left(
\begin{array}{ll}
e^{{\mbox{i}}\left( px^{+}+qx^{-}\right) } & 0 \\
0 & e^{-{\mbox{i}}\left( px^{+}+qx^{-}\right) }%
\end{array}%
\right) .  \label{dg}
\end{equation}%
The corresponding $V({\lambda })$ is
\begin{equation}
V({\lambda })=\left(
\begin{array}{ll}
\omega ({\lambda }) & 0 \\
0 & \omega ^{-1}({\lambda })%
\end{array}%
\right) \ ,  \label{phi0}
\end{equation}%
where
\begin{equation}
\omega ({\lambda })=\exp {\mbox{i}}\left( \frac{1}{1-\lambda }px^{+}+\frac{1%
}{1+\lambda }qx^{-}\right) .  \label{lmu}
\end{equation}%
In this sense $g,j_{\pm },$ and $V$ constitute the seed solution for the
Darboux transformation. Taking ${\lambda }_{1}=\mu $ and ${\lambda }_{2}=%
\bar{\mu},$ we have the following $2\times 2$ matrix solution of the Lax
pair at $\Lambda =\left(
\begin{array}{cc}
\mu & 0 \\
0 & \bar{\mu}%
\end{array}%
\right) $%
\begin{eqnarray}
M &=&\left( V(\mu )\left\vert 1\right\rangle ,V(\bar{\mu})\left\vert
2\right\rangle \right) =(\left\vert m_{1}\right\rangle ,\left\vert
m_{2}\right\rangle ),  \notag \\
&=&\left(
\begin{array}{ll}
\omega (\mu ) & \omega (\bar{\mu}) \\
-\omega ^{-1}(\mu ) & \omega ^{-1}(\bar{\mu})%
\end{array}%
\right) .  \label{hmatrix1}
\end{eqnarray}%
The reality condition (\ref{reality}) on $V$ implies that
\begin{eqnarray}
\bar{\omega}(\bar{\mu}) &=&\omega ^{-1}(\mu )\ ,  \notag \\
\omega (\mu ) &=&\bar{\omega}^{-1}(\bar{\mu}).  \label{lmu1}
\end{eqnarray}%
By direct calculations , we note that the $S$ matrix in this case is given
by
\begin{eqnarray}
S &=&{M}\Lambda {M}^{-1},  \notag \\
&=&\frac{1}{e^{r}+e^{-r}}\left(
\begin{array}{ll}
{\mu }{e^{r}}+{\bar{\mu}}{e^{-r}} & \left( {\bar{\mu}-\mu }\right) e^{{i}s}
\\
\left( {\bar{\mu}-\mu }\right) e^{-is} & {\bar{\mu}}{e^{r}}+{\mu }{e^{-r}}%
\end{array}%
\right) ,  \label{smatrix}
\end{eqnarray}%
where the functions $r(x^{+},x^{-})$ and $s(x^{+},x^{-})$ are defined by
\begin{eqnarray}
r(x^{+},x^{-}) &=&{i}\left( \frac{1}{\left( 1-\mu \right) }-\frac{1}{\left(
1-\bar{\mu}\right) }\right) px^{+}+{i}\left( \frac{1}{\left( 1+\mu \right) }-%
\frac{1}{\left( 1+\bar{\mu}\right) }\right) qx^{-},  \notag \\
s(x^{+},x^{-}) &=&\left( \frac{1}{\left( 1-\mu \right) }+\frac{1}{\left( 1-%
\bar{\mu}\right) }\right) px^{+}+\left( \frac{1}{\left( 1+\mu \right) }+%
\frac{1}{\left( 1+\bar{\mu}\right) }\right) qx^{-}.  \label{rs}
\end{eqnarray}

Let us take the eigenvalue to be $\mu =e^{i\theta }.$ The expression (\ref%
{smatrix}) then becomes%
\begin{equation}
S=\left(
\begin{array}{cc}
\cos \theta +i\sin \theta \tanh r & -i\left( \sin \theta \text{sech}r\right)
e^{is} \\
-i\left( \sin \theta \text{sech}r\right) e^{-is} & \cos \theta -i\sin \theta
\tanh r%
\end{array}%
\right) ,
\end{equation}%
and the corresponding Darboux matrix $D\left( \lambda \right) $ in this case
is
\begin{equation}
D\left( \lambda \right) =\left(
\begin{array}{cc}
\lambda -\cos \theta -i\sin \theta \tanh r & i\left( \sin \theta \text{sech}%
r\right) e^{is} \\
i\left( \sin \theta \text{sech}r\right) e^{-is} & \lambda -\cos \theta
+i\sin \theta \tanh r%
\end{array}%
\right) .
\end{equation}%
Comparing the above equation with (\ref{DP}), we find the following
expression for the projector%
\begin{equation}
P=\left(
\begin{array}{cc}
2e^{r}\text{sech}r & -2e^{is}\text{sech}r \\
-2e^{-is}\text{sech}r & 2e^{-r}\text{sech}r%
\end{array}%
\right) ,
\end{equation}%
which is same as obtained in \cite{zakha}. The solution $\widetilde{g}$ of
chiral field equations is written as
\begin{eqnarray}
\widetilde{g} &=&\left. D\left( \lambda \right) \right\vert _{\lambda
=0}g=-Sg, \\
&=&\ \left(
\begin{array}{ll}
\widetilde{X} & \widetilde{Y} \\
-\overline{\widetilde{Y}} & \overline{\widetilde{X}}%
\end{array}%
\right) g\ ,
\end{eqnarray}%
where the matrix entries are the functions
\begin{eqnarray}
\widetilde{X} &=&-\left( \cos \theta +i\sin \theta \tanh r\right) ,
\label{gama} \\
\widetilde{Y} &=&i\left( \sin \theta \text{sech}r\right) e^{is}.
\label{beta}
\end{eqnarray}%
The above expressions indicate that the functions $\widetilde{X}$ and $%
\widetilde{Y}$ have a solitonic form. Since we have%
\begin{equation}
\widetilde{j}_{\pm }=\left( I-S\right) j_{\pm }\left( I-S\right) ^{-1}
\label{js}
\end{equation}%
Using equations (\ref{abab}) and (\ref{smatrix}) in the above equation, we
get the expressions for $\widetilde{j}_{\pm }$ as
\begin{equation}
\widetilde{j}_{+}=\left(
\begin{array}{ll}
{a} & {b} \\
-{\bar{b}} & \bar{a}%
\end{array}%
\right) ,\quad \quad \quad \widetilde{j}_{-}=\left(
\begin{array}{ll}
{c} & {d} \\
-\bar{d} & \bar{c}%
\end{array}%
\right) ,  \label{abab1}
\end{equation}%
where
\begin{eqnarray}
a &=&ip\left( 1-\left( 1+\cos \theta \right) \text{sech}^{2}r\right) ,
\notag \\
b &=&-ip\left[ \left( 1+\cos \theta \right) \tanh r+i\sin \theta \right]
\left( \text{sech}r\right) e^{is},  \notag \\
c &=&iq\left( 1-\left( 1-\cos \theta \right) \text{sech}^{2}r\right) ,
\notag \\
d &=&iq\left[ \left( 1-\cos \theta \right) \tanh r-i\sin \theta \right]
\left( \text{sech}r\right) e^{is},  \notag
\end{eqnarray}%
The Equation (\ref{abab1}) shows a new solution which we have obtained by
starting from an arbitrary seed solution. By substituting above expressions
of $a,b,c,d$ in (\ref{abab1}), we see that Tr$\widetilde{j}_{+}=$Tr$\
\widetilde{j}_{-}=0.$ Therefore $\widetilde{j}_{\pm }$ satisfy the
additional constraints for $g\in SU(2)$. Consequently, when we use the above
equations in (\ref{abab1}), we get the explicit expressions of the conserved
currents (solutions) $\widetilde{j}_{\pm }$ of the chiral models by using
the Darboux transformation.

Two-soliton solution of the chiral field is obtained by the application of
two-fold Darboux transformation%
\begin{eqnarray}
\widetilde{\widetilde{X}} &=&\frac{A+B}{\sin \theta _{2}\sin \theta
_{1}\left( \sinh r_{2}\sinh r_{1}-\cos \left( s_{2}-s_{1}\right) \right)
-\left( 1-\cos \theta _{2}\cos \theta _{1}\right) \cosh r_{1}\cosh r_{2}},
\notag \\
\widetilde{\widetilde{Y}} &=&\frac{C}{2\left[ \sin \theta _{2}\sin \theta
_{1}\left( \sinh r_{2}\sinh r_{1}-\cos \left( s_{2}-s_{1}\right) \right)
-\left( 1-\cos \theta _{2}\cos \theta _{1}\right) \cosh r_{1}\cosh r_{2}%
\right] },  \label{two-soliton}
\end{eqnarray}%
where%
\begin{eqnarray*}
A &=&\cos \theta _{2}\cosh r_{2}\cosh r_{1}+i\sinh r_{2}\sinh r_{1}\left(
\sin \theta _{2}-\sin \theta _{1}\right) -i\sin \theta _{2}\sin ^{2}\theta
_{1}\sinh r_{1}\text{sech}r_{1} \\
&&-\cos \theta _{2}\left( \cos \theta _{1}\cosh r_{1}-i\sin \theta _{1}\sinh
r_{1}\right) \left( \cos \theta _{2}\cosh r_{2}+i\sin \theta _{2}\sinh
r_{2}\right) , \\
B &=&\sin \theta _{2}\sin \theta _{1}\left[ \left( \cos \theta _{1}-i\sin
\theta _{1}\tanh r_{1}\right) e^{i\left( s_{1}-s_{2}\right) }+\left( -2\cos
\theta _{2}+\cos \theta _{1}+i\sin \theta _{1}\tanh r_{1}\right) e^{-i\left(
s_{1}-s_{2}\right) }\right] , \\
C &=&-i\sin \theta _{2}\cosh r_{1}\left[ 1-\left( \cos \theta _{1}+i\sin
\theta _{1}\tanh r_{1}\right) \left( 2\cos \theta _{2}-\cos \theta
_{1}-i\sin \theta _{1}\tanh r_{1}\right) \right] e^{is_{1}} \\
&&+i\sin \theta _{1}\cosh r_{2}\left[ 1+\left( \cos \theta _{2}+i\sin \theta
_{2}\tanh r_{2}\right) \left( 2\cos \theta _{1}-\cos \theta _{2}-i\sin
\theta _{2}\tanh r_{2}\right) \right] e^{is_{2}} \\
&&+i\sin \theta _{1}\sin \theta _{2}e^{is_{1}}\left( \sin \theta _{2}\text{%
sech}r_{2}-\sin \theta _{1}\text{sech}r_{1}e^{i\left( s_{1}-s_{2}\right)
}\right) ,
\end{eqnarray*}%
and we use the notation $X\left[ 3\right] =\widetilde{\widetilde{X}}$ and $Y%
\left[ 3\right] =\widetilde{\widetilde{Y}}$. We have generated a new
solution by starting from an arbitrary seed solution. We can use the above
equations (\ref{two-soliton}) to find the expression for $S\left[ 2\right] $%
, which can be further used to obtain the explicit expressions of the
conserved currents $j_{\pm }\left[ 3\right] .$

In the asymptotic limit for $t\rightarrow \pm \infty $, we have $%
r\rightarrow \pm \infty $ and the equation (\ref{smatrix}) becomes%
\begin{equation}
\lim_{r\rightarrow \pm \infty }S=\left(
\begin{array}{cc}
\nu & 0 \\
0 & \bar{\nu}%
\end{array}%
\right) ,  \label{sneu}
\end{equation}%
where%
\begin{eqnarray}
\nu &=&\mu ,\text{ \ for }r\rightarrow +\infty  \notag \\
&=&\bar{\mu},\text{ \ for }r\rightarrow -\infty .  \label{neumeu}
\end{eqnarray}%
For $\mu =e^{i\theta }$, equation (\ref{sneu}) becomes%
\begin{equation}
\lim_{r\rightarrow \pm \infty }S=\left(
\begin{array}{cc}
e^{\pm i\theta } & 0 \\
0 & e^{\mp i\theta }%
\end{array}%
\right) ,  \label{stheta}
\end{equation}%
and the functions $\widetilde{X}$ and $\widetilde{Y}$ in the solution $%
\widetilde{g}$ of the chiral model given by equations (\ref{gama})-(\ref%
{beta}), become%
\begin{eqnarray}
\lim_{r\rightarrow \pm \infty }\widetilde{X} &=&-e^{\pm i\theta }=-\left(
\cos \theta \pm i\sin \theta \right) ,  \notag \\
\lim_{r\rightarrow \pm \infty }\widetilde{Y} &=&0.
\end{eqnarray}%
The second iteration of Darboux transformation can be used in a similar
manner and we have from equations (\ref{two-soliton})
\begin{eqnarray}
\lim_{r\rightarrow \pm \infty }\widetilde{\widetilde{X}} &=&\exp \pm i\left(
\theta _{2}+\theta _{1}\right) =\left( \cos \left( \theta _{2}+\theta
_{1}\right) \pm i\sin \left( \theta _{2}+\theta _{1}\right) \right)  \notag
\\
\lim_{r\rightarrow \pm \infty }\widetilde{\widetilde{Y}} &=&0.
\label{asympX}
\end{eqnarray}%
We see that in the asymptotic limit, we get much simpler expressions. The
equation (\ref{asympX}) gives the asymptotic behaviour of the solution $g%
\left[ 3\right] $ and it is clear from the above expression that in the
asymptotic limit $g\left[ 3\right] $ i.e. the two-soliton solution splits
into two single soliton solutions. Similarly for the $K$-th iteration of
Darboux transformation, the multisoliton solution in the asymptotic limit is
given as
\begin{equation}
\lim_{r\rightarrow \pm \infty }g\left[ K+1\right] =\lim_{r\rightarrow \pm
\infty }\left(
\begin{array}{cc}
X\left[ K+1\right] & Y\left[ K+1\right] \\
-\bar{Y}\left[ K+1\right] & \bar{X}\left[ K+1\right]%
\end{array}%
\right) g,  \label{GK1}
\end{equation}%
where%
\begin{eqnarray}
\lim_{r\rightarrow \pm \infty }X\left[ K+1\right] &=&\left( -1\right)
^{K}\exp \pm i\left( \theta _{K}+\cdots \theta _{1}\right) ,  \notag \\
&=&\left( -1\right) ^{K}\left( \cos \left( \theta _{K}+\cdots \theta
_{1}\right) \pm i\sin \left( \theta _{K}+\cdots \theta _{1}\right) \right) ,
\notag \\
\lim_{r\rightarrow \pm \infty }Y\left[ K+1\right] &=&0,  \label{XYK}
\end{eqnarray}%
which shows that the $K$-soliton solution $g\left[ K+1\right] $ of the
chiral model, splits into $K$ single solitons, where $g$ is given by
equation (\ref{dg}). Note that the $\pm $ sign appearing in the expression (%
\ref{XYK}) due to $t\rightarrow \pm \infty $, shows that there is a phase
shift in the soliton. Therefore, we see that when $t\rightarrow \pm \infty $%
, the asymptotic solution split up into $K$ single solitons.

From the above calculations we see that, in the asymptotic limit $S\left[ k%
\right] =M\left[ k\right] \Lambda _{k}M\left[ k\right] ^{-1}\rightarrow
M_{k}\Lambda _{k}M_{k}^{-1}$. Therefore in the asymptotic limit, the
quasideterminant (\ref{gk3}) splits into $K$ factors i.e.%
\begin{eqnarray}
\lim_{r\rightarrow \pm \infty }g\left[ K+1\right] &=&\lim_{r\rightarrow \pm
\infty }\left\vert
\begin{array}{ccccc}
M_{1} & M_{2} & \cdots & M_{K} & I \\
M_{1}\Lambda _{1} & M_{2}\Lambda _{2} & \cdots & M_{K}\Lambda _{K} & O \\
M_{1}\Lambda _{1}^{2} & M_{2}\Lambda _{2}^{2} & \cdots & M_{K}\Lambda
_{K}^{2} & O \\
\vdots & \vdots & \cdots & \vdots & \vdots \\
M_{1}\Lambda _{1}^{K} & M_{2}\Lambda _{2}^{K} & \cdots & M_{K}\Lambda
_{K}^{K} & \frame{\fbox{$O$}}%
\end{array}%
\right\vert g,  \notag \\
&=&\left\vert
\begin{array}{cc}
M_{K} & I \\
M_{K}\Lambda _{K} & \frame{\fbox{$O$}}%
\end{array}%
\right\vert \left\vert
\begin{array}{cc}
M_{K-1} & I \\
M_{K-1}\Lambda _{K-1} & \frame{\fbox{$O$}}%
\end{array}%
\right\vert \cdots \left\vert
\begin{array}{cc}
M_{1} & I \\
M_{1}\Lambda _{1} & \frame{\fbox{$O$}}%
\end{array}%
\right\vert g,  \notag \\
&=&\prod\limits_{k=1}^{K}\left( -1\right) ^{k}\left\vert
\begin{array}{cc}
M_{K-k+1} & I \\
M_{K-k+1}\Lambda _{K-k+1} & \frame{\fbox{$O$}}%
\end{array}%
\right\vert g.  \label{asympg}
\end{eqnarray}%
We can say that the splitting of $K$-soliton solution into $K$ single
soliton solutions asymptotically, is in fact equivalent to the factorization
of quasideterminant solution (\ref{gk3}) into a product of quasideterminants
of $2\times 2$ matrices over a noncommutative ring $R$ of $N\times N$
matrices.

\section{Concluding remarks}

In this paper, we have considered the principal chiral model in two
dimensions, based on some Lie group and presented the quasideterminant
solutions of the chiral model as well as its Lax pair obtained by means of
Darboux transformation, defined in terms of Darboux matrix. We iterated the
Darboux transformation to get the quasideterminant multisoliton solutions.
We have also discussed the relation of the Darboux matrix approach with the
Zakharov-Mikhailov's dressing method, where the Darboux matrix was shown to
be expressed in terms of hermitian projector defined in terms of particular
column solutions of the Lax pair. At the end we calculated the one and two
soliton solutions for the case of Lie group $SU\left( 2\right) $. The
asymptotic limit of the solutions in $SU\left( 2\right) $ case splits the
solution in product of single solitons. We have also obtained the asymptotic
solution of the chiral model in terms of quasideterminant for the case of
Lie group $SU\left( 2\right) $. It would be interesting to study the
quasideterminant solutions of the supersymmetric chiral models and those of
the nonlinear sigma models based on symmetric spaces. We shall address these
issues in a separate work.

\bigskip

{\large \textbf{Acknowledgements}}

\noindent BH gratefully acknowledges the Higher Education Commission of
Pakistan for financial support through indigenous scholarship scheme for PhD
studies. MH wishes to thank Jonathan Nimmo for helpful discussions.

\bigskip

\end{document}